\documentclass[10pt,journal,compsoc]{IEEEtran} 

\ifCLASSOPTIONcompsoc
  \usepackage[nocompress]{cite}
\else
  \usepackage{cite}
\fi

\usepackage[caption=false,font=footnotesize,labelfont=sf,textfont=sf]{subfig}
\usepackage[pdftex]{graphicx} 
\DeclareGraphicsExtensions{.jpg, .pdf}
\usepackage{tabularx}

\begin{document}

\title{
A Survey of Storage Class Memory: \\Principles, Problems, and Possibilities
}

\author{Aditya~K~Kamath,
        Leslie~Monis,
        A~Tarun~Karthik,
        and~Basavaraj~Talawar
\IEEEcompsocitemizethanks{\IEEEcompsocthanksitem A. Kamath, L. Monis, A. Karthik, and B. Talawar are with the Department of Computer Science and Engineering, National Institute of Technology Karnataka, Surathkal,
Karnataka, 575025.\protect\\
E-mails: akamath1997@gmail.com, lesliemonis@gmail.com, tarunkarthik999@gmail.com, basavaraj@nitk.edu.in}
\thanks{Manuscript received -; revised -.}}

\markboth{}%
{A. Kamath \MakeLowercase{\textit{et al.}}: A Survey of Storage Class Memory}

\IEEEtitleabstractindextext{%
\begin{abstract}

Storage Class Memory (SCM) is a class of memory technology which has recently become viable for use. Their name arises from the fact that they exhibit non-volatility of data, similar to secondary storage while also having latencies comparable to primary memory and byte-addressibility. In this area, Phase Change Memory (PCM), Spin-Transfer-Torque Random Access Memory (STT-RAM), and Resistive RAM (ReRAM) have emerged as the major contenders for commercial and industrial use. In this paper, we describe how these memory types function, while highlighting the problems of endurance and performance that these memory types face. We also discuss the future possibilities of Multi-Level Cells (MLCs), as well as how SCM can be used to construct accelerators.

\end{abstract}

}


\maketitle

\IEEEpeerreviewmaketitle

\section{Introduction} \label{Introduction}
\IEEEPARstart{N}{on-Volatile} Memory (NVM) is a special class of memory that exhibits persistence, similar to that of secondary memory, while providing access speeds at least two magnitudes faster. NVM can potentially replace or augment any of the current existing memory layers, like that of the cache or primary memory. Storage Class Memory (SCM) is a subset of NVM whereby the device exhibits the feature of data persistence, while offering performance comparable to or better than that of primary memory along with byte-addressibility \cite{5388605, Hameed:2017:ESL:3132402.3132414, 7516054}.

The current technology used for designing caches and primary memory face many problems that SCM could potentially solve. Static RAM (SRAM), the technology typically used in caches, suffers from low density making it increasingly difficult to pack together in order to meet the growing demands of speed \cite{MITTAL201433}. Dynamic RAM (DRAM) used in primary memory has better density, but suffers from slower access times, and requires constant power to refresh memory \cite{Mittal:2012:SAT:2421508.2421513}.

Three emerging SCM technologies are Phase Change Memory (PCM), Spin-Transfer-Torque Random Access Memory (STT-RAM), and Resistive Random Access Memory (ReRAM). As Table \ref{tbl:device_properties} and Table \ref{tbl:cache_comparison} indicate, their main advantages are that they exhibit low leakage power dissipation, since they do not need to be refreshed constantly. They also have the ability to be packed close together leading to higher data storage for the same given volume when compared to SRAM and DRAM. This is since the cell size of STT-RAM and ReRAM, considered as possible replacements for SRAM, are nearly one-tenth the size of SRAM. Another important feature that these memory devices exhibit is their ability to function as Multi-Level Cells (MLCs) \cite{jlpea7030023}, this is where a single memory cell is capable of storing more than one bit. As the resistance of the memory element can be varied based on current supplied for SCM technology, MLC is achieved by assigning different resistance levels different bit values. 

While these all seem like promising reasons to adopt SCM technology, they also face some severe drawbacks. PCM and STT-RAM both suffer from high write latencies \cite{7120149}, nearly ten times that of DRAM for PCM, and ten times SRAM for STT-RAM, and high write energy. PCM and ReRAM also suffer from a limit on write endurance before a hard error occurs. Unlike soft errors, a hard error is when a memory element gets permanently stuck at a certain value and cannot be changed \cite{Schechter:2010:UEE:1816038.1815980}. These are major issues that needs to be tackled before SCM can be widely adopted.

In this paper, we give a brief overview of how the different types of SCM function. We then identify the key areas of focus, classifying the research being done in these fields.

Section \ref{SCM} covers the details of how the three SCM based memories PCM, STT-RAM, and ReRAM function. Section \ref{Research} covers all the recent research trends that have been on-going in this field. Section \ref{Simulators} gives a brief introduction of popular simulators oriented towards studying properties of SCM memory. Section \ref{Future} gives an overview of possibilities that may emerge in future.

\section{SCM Characteristics} \label{SCM}
\subsection{Phase Change Memory (PCM)} \label{PCM}
\begin{figure}[h]
    \centering
    \subfloat[]{\includegraphics[width=2.5in]{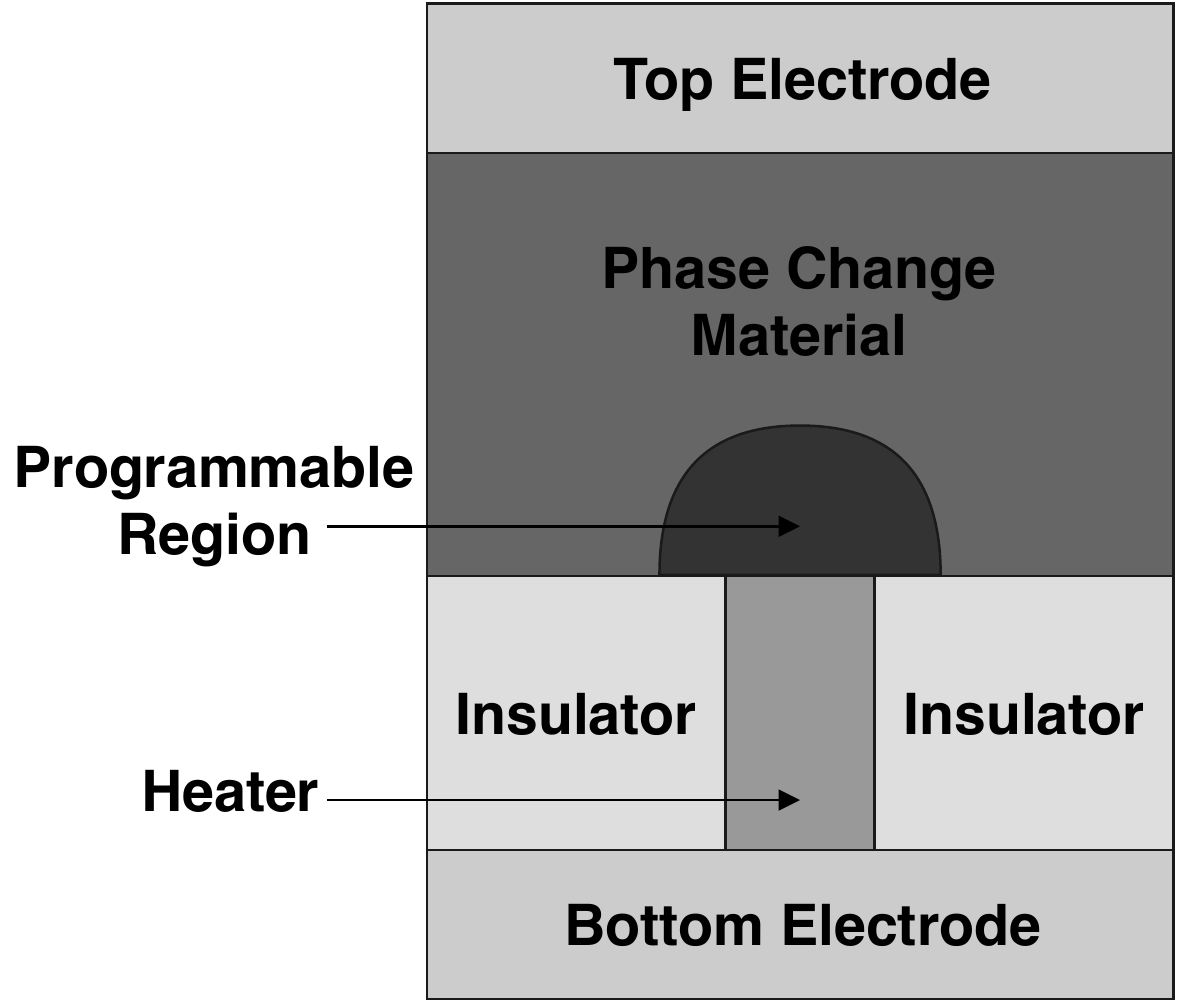}\label{fig:PCM_Structure}}
    \hfil
    \subfloat[]{\includegraphics[width=2.5in]{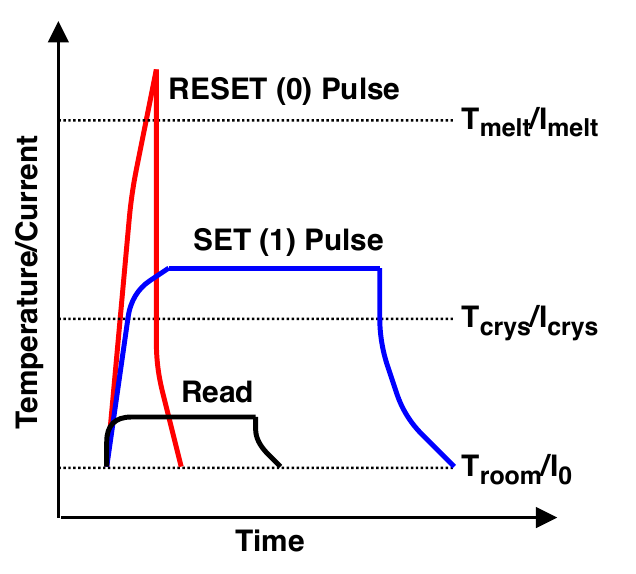}\label{fig:PCM_Graph}}
    \caption{(a) Cross-section of PCM cell. (b) Temperature/Current and Time required to program or read PCM cell. \cite{5609179} }
    \label{fig:PCM}
\end{figure}

Figure \ref{fig:PCM} shows the basic view of a PCM cell, along with the requirements for programming the cell. The phase change material is a chalcogenide typically consisting of Ge-Sb-Te (GST) \cite{hudgens_johnson_2004}. GST can exist in two states, amorphous and crystalline. In the amorphous state, the device exhibits higher resistance than the crystalline state. These two states can be used to store data, where high resistance is assigned a value of 0, and low resistance is assigned a 1. PCM can also be used to store multiple states as a Multi-Level Cell (MLC). This is by assigning data values to intermediate resistances. In order to SET the cell, it must reach a temperature below melting point, but above crystallization temperature. The cell needs time to change into crystalline state, which is why the pulse lasts much longer. For RESET, the cell temperature must be rapidly increased to melting point, then quickly cooled so that the material sets into amorphous state \cite{5388621, 1269376}. 

\subsection{Spin-Transfer Torque Random Access Memory (STT-RAM)} \label{STT}
\begin{figure}[h]
    \centering
    \subfloat[]{\includegraphics[width=1.25in]{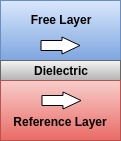}\label{fig:STT-RAM-SET}}
    \hfil
    \subfloat[]{\includegraphics[width=1.25in]{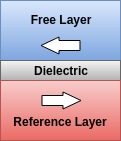}\label{fig:STT-RAM-RESET}}
    \hfil
    \caption{(a) STT-RAM cell in SET (1) state. (b) STT-RAM cell in RESET (0) state.}
    \label{fig:STT-RAM}
\end{figure}

STT-RAM represents bits of data by relying on differences in magnetic directions. There are two ferromagnetic layers in an STT-RAM separated by a dielectric. One layer is referred to as the reference layer, as it has a fixed magnetization direction. The other layer is referred to as free layer, whose magnetization direction can be controlled by passing current. Based on the relative directions of the two layers, the resistance of the magnetic tunnel junction (MTJ), will differ. In the case where the magnetization direction of the two layers are aligned, the resistance at the MTJ will be low, indicating a state of 1. If the two layers have opposing directions, the resistance becomes high, indicating a state of 0 \cite{5993623}. Figure \ref{fig:STT-RAM} shows the arrangement of the cell for the two states.

\subsection{Resistive Random Access Memory (ReRAM)} \label{ReRAM}
Resistive RAM experiences resistance changes due to electrochemical effects. The ReRAM cell consists of two metal electrodes, separated by a metal oxide layer. The behaviour of this system is dependent on the concentration of the oxygen vacancy in the metal oxide layer. By applying current to the cell, the state the cell is in can be switched. In the case of a bipolar ReRAM cell, the SET operation is undertaken when a negative bias is applied, while the RESET operation is undertaken when a positive bias is applied. An example of a ReRAM cell is a titanium oxide layer sandwiched between two platinum electrodes \cite{JOUR}.

\begin{figure}[h]
    \centering
    \subfloat[]{\includegraphics[width=1.5in]{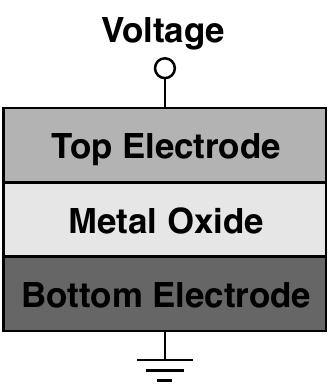}\label{fig:ReRAM_Structure}}
    \hfil
    \subfloat[]{\includegraphics[width=1.5in]{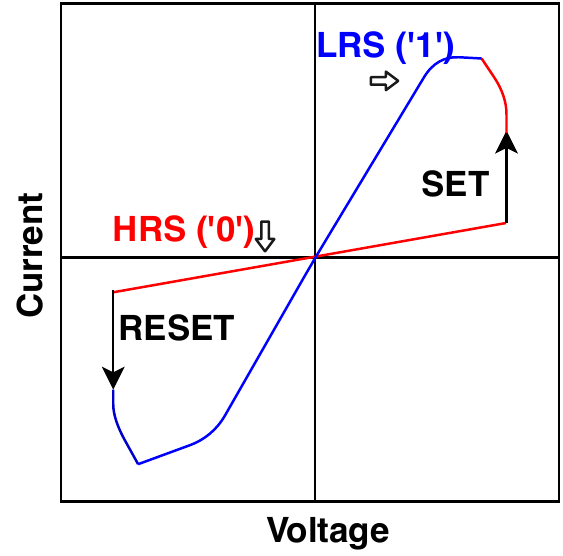}\label{fig:ReRAM_Graph}}
    \caption{(a) Structure of ReRAM cell. (b) Bias and current for write operations. \cite{5609179}}
    \label{fig:ReRAM}
\end{figure}

\begin{table*}
\caption{Comparison of Device Properties of Memory Technologies \cite{7120149}}
\label{tbl:device_properties}
\begin{center}
\begin{tabular}{|c||c|c|c|c|c|c|c|}
\hline
 & Cell size $(F^2)$ & Access Granularity & Read Latency & Write Latency & Erase Latency & Endurance & Standby Power\\
\hline
HDD & N/A & 512B & 5 ms & 5 ms & N/A & $\geq 10^{15}$ & 1 W\\
\hline
SLC Flash & 4 - 6 & 4KB & $25 \mu s$ & $500 \mu s$ & 2 ms & $10^4 - 10^5$ & 0\\
\hline
DRAM & 6 - 10 & 64B & 50 ns & 50 ns & N/A & $\geq 10^{15}$ & Refresh Power\\
\hline
PCM & 4 - 12 & 64B & 50 ns & 500 ns & N/A & $10^8 - 10^9$ & 0\\
\hline
STT-RAM & 6 - 50 & 64B & 10 ns & 50 ns & N/A & $\geq 10^{15}$ & 0\\
\hline
ReRAM & 4 - 10 & 64B & 10 ns & 50 ns & N/A & $10^{11}$ & 0\\
\hline
\end{tabular}
\end{center}
\end{table*}

\begin{table}
\caption{Comparison of Cache Memory Technologies \cite{6816046}}
\label{tbl:cache_comparison}
\begin{tabular}{|c||c|c|c|c|}
\hline
 & SRAM & DRAM & STT-RAM & PCM\\
\hline
Cell Size $(F^2)$ & 120 - 200 & 4 - 6 & 6 - 50 & 4 - 12\\
\hline
Multi-level cell & No & No & Yes & Yes\\
\hline
Read speed & Very fast & Slow & Fast & Slow\\
\hline
Write speed & Very fast & Slow & Slow & Very slow\\
\hline
Read energy & Low & Medium & Low & Medium\\
\hline
Write energy & Low & Medium & High & High\\
\hline
Leakage & High & Medium & Low & Low\\
\hline
Throughput & Very high & Medium & High & Low\\
\hline
Write Endurance & $10^{16}$ & $10^{16}$ & $\geq 10^{12}$ & $10^{8} - 10^{9}$\\
\hline
Soft Error & Low & High & No & No\\
\hline
\end{tabular}
\end{table}

\section{Current Research Areas} \label{Research}
\begin{figure}[h]
    \centering
    \includegraphics[width=3.5in]{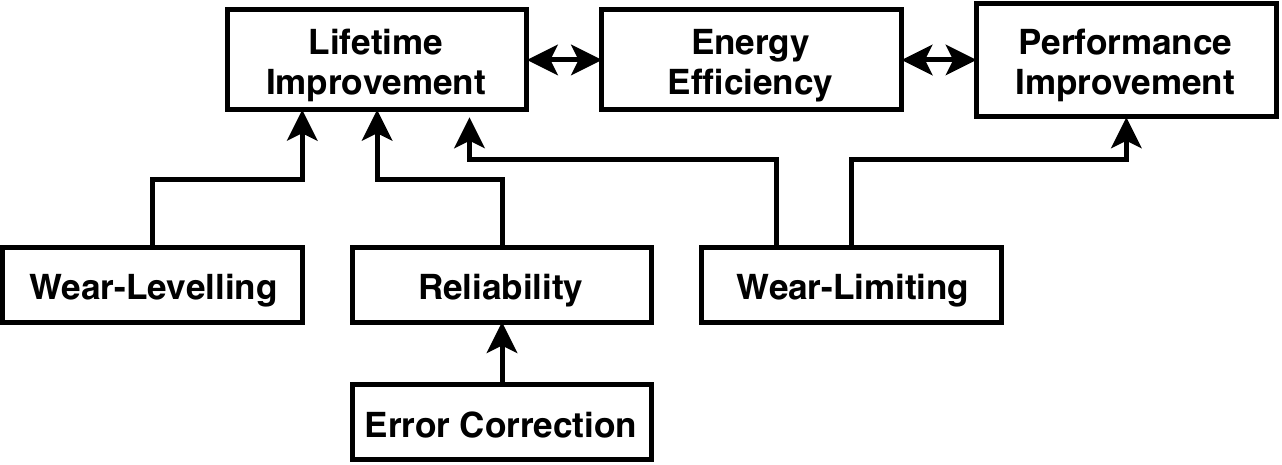}
    \hfil
    \caption{Relations between different SCM issues. An arrow from one topic to another implies that changes in that field may cause impacts in the other topic.}
    \label{fig:research}
\end{figure}

The majority of SCM is still in the experimental stage and there is a lot of ongoing research on possible outcomes of these memory devices. This section lists several research areas which are currently popular in this field. Table \ref{tbl:research} lists all the areas along with the relevant papers. Some of these topics have relations where an impact in one topic may end up causing an impact in another topic. Thus, certain papers end up accomplishing multiple objectives and have been repeated multiple times in the table. Figure \ref{fig:research} shows some of the relations between solutions to problems that SCM faces. 

\subsection{Lifetime Improvement}
While PCM has many advantages over DRAM memory, including non-volatility, low standby power, and high density, a major cause of concern is the endurance \cite{5388621}. PCM and ReRAM technologies have a write endurance that are magnitudes less than those of memory technologies currently in use. Due to this, the lifetime of these memories is severely hindered. 
\subsubsection{Wear-leveling}
These are techniques which attempt to distribute writes evenly over all cells, by continuously trying to change the cell that write operations take place. Wear-leveling techniques already exist \cite{ban2004wear, kgil2008improving} and are used for NAND Flash based SSDs. These involve creating a logical to physical mapping of addresses, storing the number of writes a line experiences in a table and using that data to periodically change the mapping. A shortcoming of this is that it requires a high overhead for the tables, and increases latencies due to reading and resolving of mappings. Qureshi et al. \cite{Qureshi:2009:ELS:1669112.1669117} proposed an alternative system which uses address space randomization to provide a low overhead wear-leveling approach. 

Caches are optimised in order to maximise the temporal locality of the data in order to improve performance. This can lead to a disproportionate amount of writes being directed to certain cache lines \cite{Qureshi:2009:ELS:1669112.1669117}. Endurance problems can thus arise when these caches consist of SCM. When the lifetime of the cache is estimated, the assumption is that writes are distributed evenly across the cache. Due to the disproportionate writes, certain cache lines may start to fail much before the estimated lifetime. Wear-leveling techniques are thus required in order to avoid this situation. Two simple types of wear leveling schemes arise for caches: the first is intra-set wear-leveling, where an attempt is made to distribute writes evenly within a cache set; the second scheme is inter-set wear-leveling, where an attempt is made to avoid one cache set from receiving more writes than the other sets. In this way the writes are distributed across the entire cache. For the most part, both of these schemes can operate independently, allowing for one inter-set and one intra-set wear-leveling methodology to coexist in the same cache.

One inter-set wear-leveling scheme involves set remapping \cite{Chen:2013:OCB:2463585.2463592} which has been proposed to try to tackle this problem. In this proposal, a register is maintained to determine which set to write to. After a certain interval of time the value of the register is changed, thus changing the location that data is written to. This method requires extra tags to maintain set index, increasing the memory overhead. Wang et al. \cite{6522322} tried to introduce inter-set and intra-set wear-leveling using two methods. For inter-set leveling, the number of writes are measured, and after a threshold is reached, two sets are swapped. Data that was present in the sets before the swap are then invalidated. Since only two sets are swapped at a time, the performance does not take a severe toll. For intra-set leveling, the number of write hits is kept track of using a global counter. When the counter saturates, the cache line last written is flushed from the cache. This can cause a decrease in performance, and does not guarantee that the cache line that was flushed was actually frequently used. 

Another proposal \cite{Mittal:2014:WIL:2591513.2591525} breaks a cache set into multiple modules. The amount of writes each module receives is noted, and if the variation reaches a certain threshold, data in the most written module is moved to the least written module, and the most written module is temporarily disabled. In this way intra-set wear-leveling is achieved. A limitation is that this requires complicated computational circuitry when calculating the variation. Similar to this, Agarwal et al. \cite{8611205} proposed partitioning the cache into multiple windows. For each window, the number of writes received is noted. At regular intervals, the window with the most writes is set to read-only, and the counter for that window is reset achieveing intra-set wear-leveling. Mittal et al. \cite{7027229} proposed an intra-set wear-leveling mechanism which required maintaining a counter for each cache line. Comparisons are made between the counters of the same set, and when the difference between the largest and smallest value crosses a threshold, their data is swapped, and their counters are reset. The larger the associativity of the cache, the better the performance in this scheme.  
\subsubsection{Wear-limiting}
These are techniques which attempt to reduce the overall number of writes required for functioning. Data Comparison Write \cite{Zhou:2009:DEE:1555815.1555759} schemes achieve this by first issuing a read on a cell, and then only writing to the cell if the written value differs from the stored value. Flip-N-Write \cite{5375405} is another technique which builds on this, to reduce the number of writes. This is done by allocating an extra bit per block. When a write is issued to a block, the number of bit write operations needed is checked. If it is greater than half the size of the block, the extra bit is set, and each bit of the data is complemented, then stored. Otherwise a normal data comparison write takes place. This method guarantees that a block write will require at maximum, bit writes equal to half the size of the block.

While wear-leveling and wear-limiting techniques help reduce the impact of writes, they do not have any effect on the limit on the number of writes that a cell can experience. An analytical model \cite{Strukov2016} found that there was a linear to cubic tradeoff between endurance and write speed. Zhang et al. \cite{Zhang:2016:MWE:3001136.3001192} built on this, proposing a wear-limiting methodology to reduce the wear that a single write operation causes by performing slow writes. Three different mechanisms, Bank-Aware, Eager and Wear Quota, were discussed in order to slow writes without adversely affecting the performance. Results show that this could potentially result in the lifetime doubling.
\subsection{Error-Correction}
Recent research \cite{Sridharan:2015:MEM:2786763.2694348, 7266869} has shown that a common cause of system crashing is due to memory failures. While many methods have arisen that attempt to reduce the effect of writes on cells and thus increase the lifetime of SCM, hard errors may still occur in the memory. By improving the error correction capabilities of SCM, the lifetime of the memory will improve, as the system will be able to last even after errors start occurring. Due to this, some procedures have to be developed in order to overcome these errors when they arise. Error-Correcting Codes (ECCs) \cite{5390376, 5763257} have existed to mitigate effects of soft errors that occur in DRAM and SRAM. However, these approaches cannot be applied directly for SCM, as they have a high memory overhead. New methods for error mitigation have to be explored. 

\begin{figure}[h]
    \centering
    \includegraphics[width=3.5in]{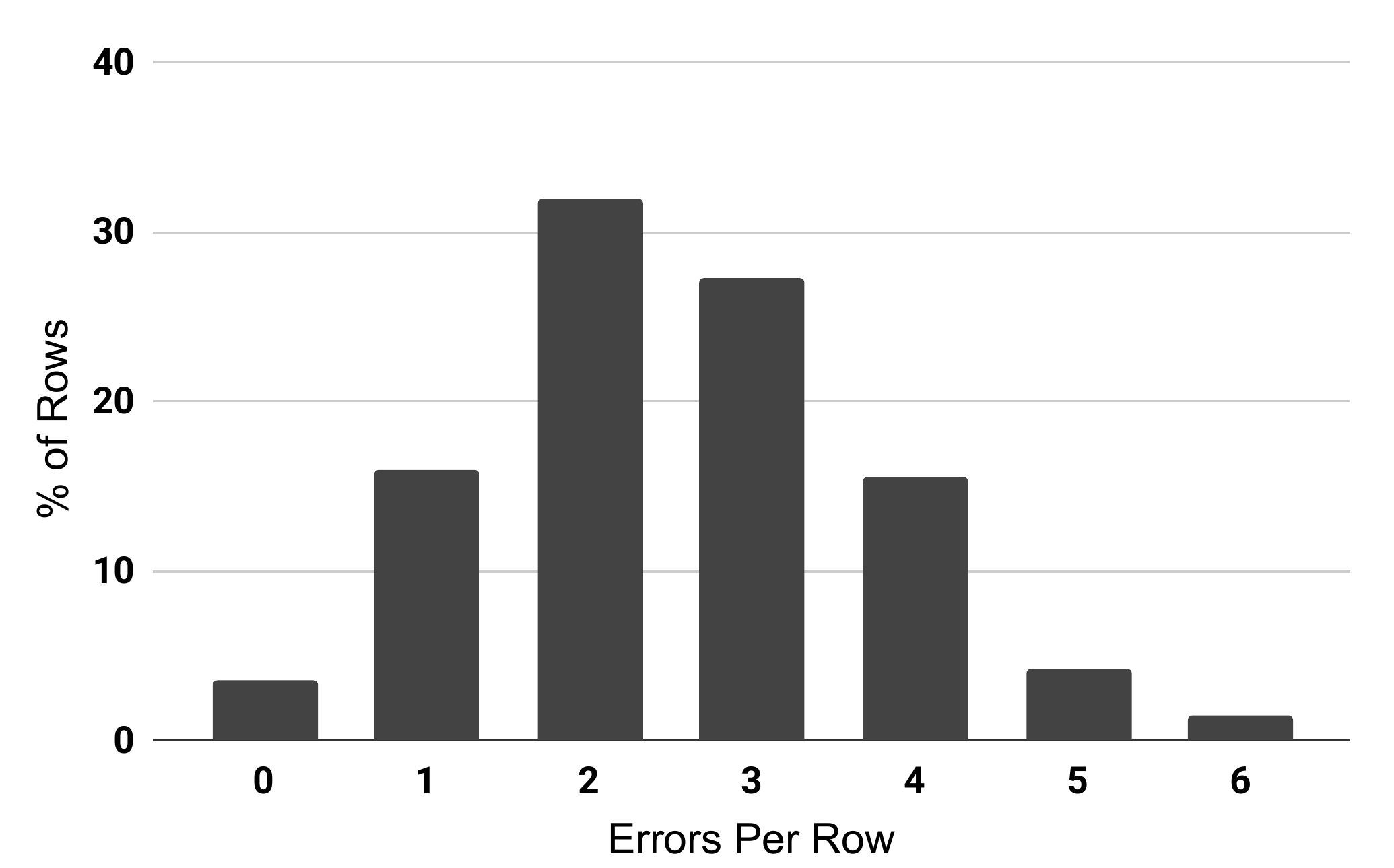}
    \hfil
    \caption{Distribution of number of errors in a row when a page reaches failure point. \cite{Swami:2017:EES:3154814.3151083}}
    \label{fig:error_correction}
\end{figure}

Error-Correcting Pointers (ECPs) \cite{Schechter:2010:UEE:1816038.1815980} were suggested as an alternative for ECCs in SCM. An ECP is a pointer to a failed cell, which also stores the correct value for that cell. The suggestion given was to provide 6 ECPs for every 512-bit memory block. This enables each 512 bit block to recover from at most 6 hard errors. One limitation is that uniform allocation is not optimal. Certain rows may be more likely to experience hard failures, while others may be less likely. Once a 512 bit block fails beyond recovery, the entire page has to be discarded. As we can see from Figure \ref{fig:error_correction} the vast majority of rows do not even reach four errors when the failure point of the page is reached. To counter this, proposals \cite{7851482, Azevedo:2013:ZME:2508148.2485961} have been made to improve usage of ECPs, at the cost of additional latency and hardware. Other methods \cite{5749752, Ipek:2010:DRM:1735971.1736023} rely on the operating system for providing support.

Use of Error-Correcting Strings (ECSs) \cite{Swami:2017:EES:3154814.3151083} is another proposal, in which variable-length offsets are used instead of fixed-length pointers. This allows for larger tolerance of hard errors until failure occurs. This is combined with a page-level error correction, which provides ECSs to blocks on demand. To reduce the wastage of pages caused by failures, Tavana et al. \cite{Tavana:2018:BCA:3274266.3243906} suggest utilizing Aegis \cite{7847645} alongside ECP for error correction, and suggesting a method of block cooperation where the unused metadata space of one block is shared with other nearby blocks in order to reduce the chance of failure of the entire page. 
 
\subsection{Performance and Parallelism}
The previously mentioned technique, Flip-N-Write \cite{5375405}, was a major milestone in the improvement of write performance. It provided a strict upper-bound on the time that would be required to complete a write operation on a block of data. It also reduced the worst-case number of write operations in half. Recent works are looking at further improving the write performance of SCM by writing multiple cells in parallel. For clarity, a \textit{bit write} refers to a single bit whose value is changing, while a \textit{write operation} refers to a single write issued to the controller which may consist of multiple bit writes.

\begin{figure*}[h]
    \centering
    \includegraphics[width=7in]{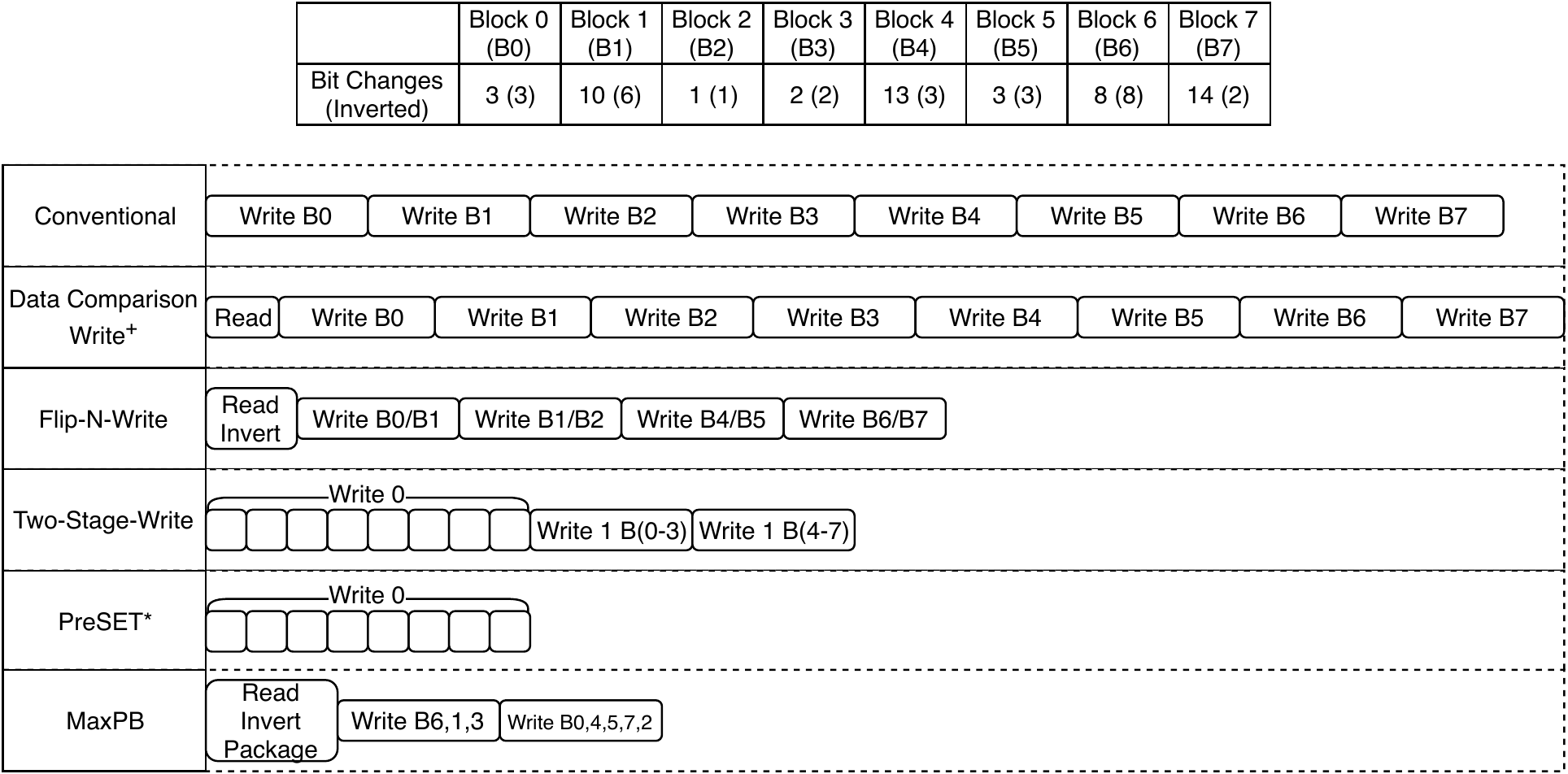}
    \hfil
    \caption{Rough comparison of time taken for Conventional, Data Comparison Write \cite{Zhou:2009:DEE:1555815.1555759}, Flip-N-Write \cite{5375405}, Two-Stage-Write \cite{6522326}, PreSET \cite{Qureshi:2012:PIP:2366231.2337203}, and MaxPB \cite{Li:2016:MAP:3012405.3012007} write schemes, assuming blocks of 16 bits are written at a time. ($^+$) While Data Comparison Write has the overall highest latency, it saves both energy and endurance over the Conventional scheme, by first reading the cell. ($^*$) While PreSET has the overall least latency, it reduces the lifetime of the memory, due to writing SET regardless of whether it's required.}
    \label{fig:parallelism}
\end{figure*}

Many techniques try to improve write parallelism by taking advantage of the asymmetries that exist in PCM. Writing a 0 (RESET) in PCM takes more current, but requires less time compared to writing a 1 (SET). Qureshi et al. \cite{Qureshi:2012:PIP:2366231.2337203} proposed a PreSET mechanism where all the cells of a PCM main memory line are preemptively SET when the respective cache line becomes dirty. In this way, when a writeback request is sent to main memory, only the faster RESET operations need to take place. This mechanism incurs a large lifetime penalty, as many unnecessary SET operations are required. Another method referred to as two-stage-write \cite{6522326}  exploits the write asymmetries by breaking a write operation into two stages. In the first stage, bits that are 0 are written at an accelerated rate. This is followed by the second stage, where bits that are 1 are written into in parallel to the extent that the power constraints allow. While this method improves write performance, it requires significant write asymmetries to give viable improvements. In addition, due to the separate stages, extra control circuitry is required. Li et al. \cite{7573814} propose an approach where the number of 1s and 0s to be written are measured. Following this, a schedule is created keeping in mind the asymmetries of power and time. In this way, concurrent bit writes are arranged without violating the power requirements. An alternative proposal called MaxPB \cite{Li:2016:MAP:3012405.3012007} involves estimating the power requirements of each write operation that needs to be undertaken. Based on this, and the power budget given, the write operations are packaged together in order to obtain the minimum number of write operations required. The write operations in a single package are then executed in parallel, allowing for an improvement in speed. Figure \ref{fig:parallelism} compares the time taken for a few of the previously mentioned schemes.

\begin{figure}[h]
    \centering
    \includegraphics[width=3.5in]{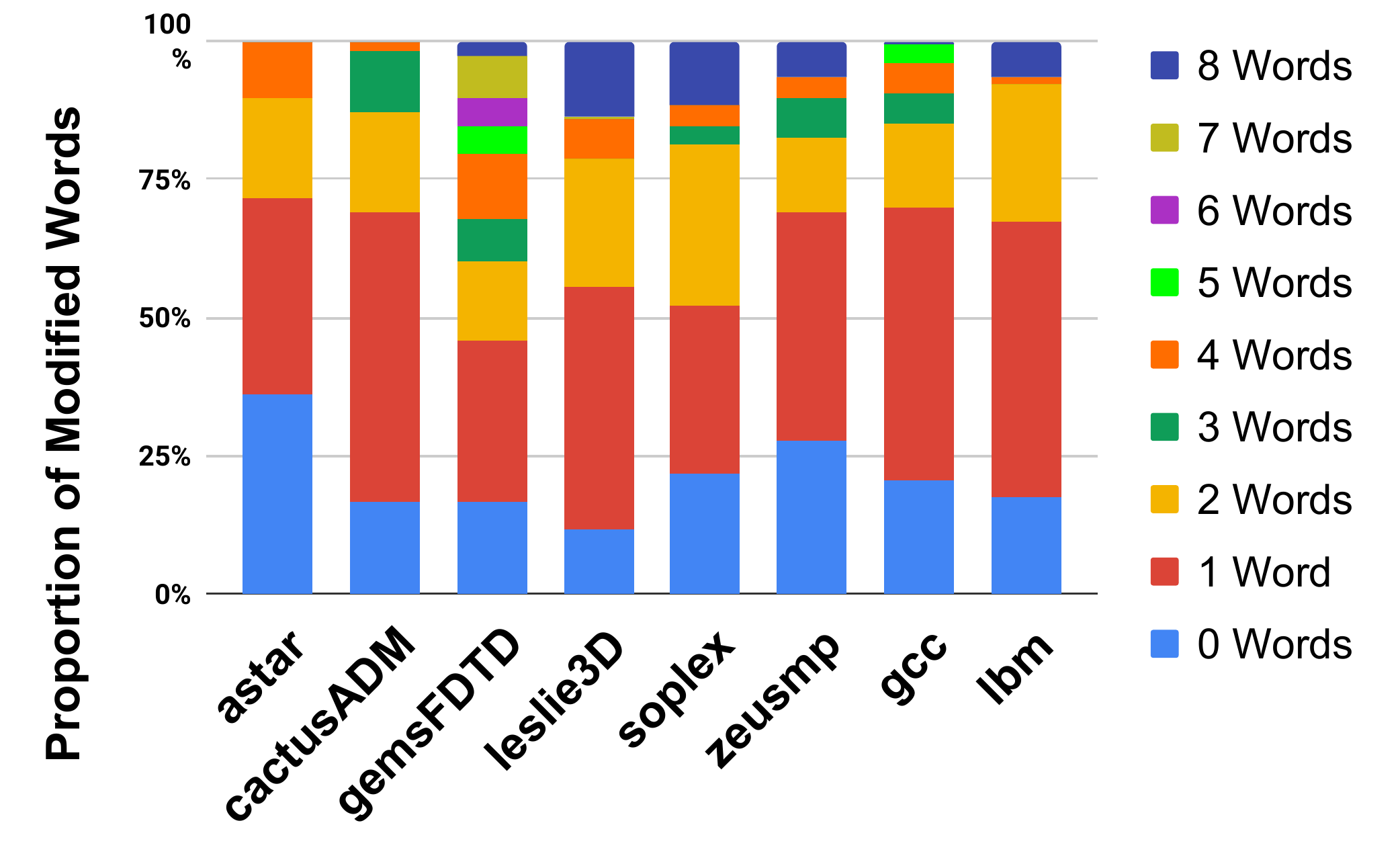}
    \hfil
    \caption{Percentage of cache lines writes with given number of modified words, assuming a cache line is 64 bytes and a word is 8 bytes. \cite{7551433}}
    \label{fig:words}
\end{figure}

An issue faced by writes on PCM memory is that write operations are handled at a cache line granularity. However, the entire cache line may not have been modified, or the write may in fact be a Silent Store \cite{898055}. Figure \ref{fig:words} shows the percentage of cache lines with the number of modified words for a subset of the SPEC CPU 2006 \cite{Spradling:2007:SCB:1241601.1241625} benchmark suite. 0 words modified means that the write was a Silent Store \cite{898055}, while 8 means that the entire cache line was modified. It can be clearly seen that in the majority of cases, only a small part of the cache line has actually been modified. Another problem arising is that while the write is being handled, the unmodified chips within a rank will remain idle. A proposal \cite{Xia:2014:DDW:2597652.2597661} was made to coalesce writes that effect the same rank into a single write operation. In this way, multiple write operations that target the same rank can be sent as a single operation, reducing the latencies and energy requirements that the multiple operations would have consumed. Similarily, MaxPB addressed this issue in order to reduce the time that write operations took. Since read operations are on the critical path, writes that are issued are put into a buffer until a certain threshold is reached. Once this threshold is reached, write operations that are required are undertaken until a lower threshold is reached. In this way, reads and writes occur in bursts. PCM also has a smaller write bandwidth than DRAM \cite{6176872, MICRON}. Due to this, while write operations are undertaken, read operations are forced to wait, increasing their overall latency. One proposal \cite{5416645} made to address this, was to cancel or pause writes when read operations were necessary. Arjomand et al. \cite{7551433} proposed several mechanisms involving utilizing error correction codes and rotating data mappings to allow for parallel read and write operations for chips that are currently being written into.

\subsection{Multi-Level Cell}
The resistance range of the cells in PCM are fairly large, due to the resistance gap between the amorphous and crystalline states. By manipulating the temperature and duration of a write operation to PCM, it is found that resistance values intermediate to the two states can be achieved. This allows for the possibility of a single cell to hold multiple bits by using intermediate resistance values for different states. These cells are thus referred to as Multi-Level Cells (MLCs). By using MLCs instead of the previous Single-Level Cells (SLCs), the density of PCM is greatly increased. In addition, the extra storage that is gained can help in tolerating failures. 

However, MLCs comes with their own set of problems. Due to smaller gaps in resistance between bit levels, the write operation is more complicated and must be more precise. It is performed by applying multiple iterations of Program and Verify (P\&V) \cite{4418973, 6169027}. The operation takes place because the exact parameters (current/temperature) with which to issue the write to achieve a certain resistance varies between cells. Thus, an estimation is made and a write operation is undertaken. The value is then read, and if the resistance is not within the desired range, the process is repeated. This results in latencies almost 4 to 8 times higher than in SLC \cite{5416645}. In addition, the write energy consumption is also increased. MLCs also experience problems from resistance drift \cite{6241871, 5958219}. Resistance drift is a phenomenon where the resistance of PCM cells increase with time. In an SLC, since the resistance gap was so large this issue did not cause any problems, but in an MLC the gaps are smaller and resistance drift can cause an unwanted change in state of the cell. This causes soft errors to pop up in PCM cells, and reduces the reliability of the device.

To resolve issues with reliability, the concept of guardbands are used \cite{4418973}. In this, a band of resistance is kept between consecutive states. When a P\&V operation takes place, it is ensured that the final value does not lie within the guardband range. The size of the guardband determines the amount of resistance drift that can be endured. A larger guardband means that data is retained longer as a larger resistance drift can be endured, at the cost of a larger number of iterations of P\&V. A smaller guardband means that writes are less costly in terms of latency and energy, but the data will not be retained as long. There is thus a clear latency/retention time tradeoff involved \cite{Li:2013:CDW:2499369.2465564}.  It has been shown that the number of P\&V iterations undertaken have no effect on the endurance of the cell \cite{1493077}, because RESET pulses determine endurance more than SET. To take advantage of all of these functionalities, Zhang et al. \cite{7920841} proposed that two types of write operations exist. One, a long latency, high retention write, and the second a low latency, low retention write. They proposed a region retention monitor, to guess the required type of write when write operations are issued. Based on this guess, the appropriate write operation takes place. In this way, latencies can be improved for data that does not require high retention time.

Another method \cite{Seong:2013:TPC:2485922.2485960} proposed to overcome the reliability problems was to reduce the number of possible states that each cell of MLC has from 4 to 3, combining the two states with small resistance gap into a single state. Rashidi et al. \cite{Rashidi:2018:IMP:3199680.3177965} combined this with utilising the extra storage reserved for hard error correction to further alleviate the issue of reliability. This extra storage typically remains under-utilized for a long time, until hard errors start to occur. M-Metric \cite{doi:10.1063/1.3653279} alongside this extra error-correction metadata is used with a tri-level PCM to improve the latencies and energy consumption as well as increase the overall IPC. 

Compression techniques have been proposed as a method of reducing the impact on energy, lifetime, and latencies of MLC SCMs. A few techniques \cite{6880482, Sardashti:2013:DCC:2540708.2540715} leverage compression to try to reduce the number of individual write operations required. However, these methods have a high overhead in terms of memory and computation. Another method is CompEx \cite{7446056} coding, which attempts to provide a low overhead solution to classical methods of compression by combing statistical compression with expansion coding. Building on this, CompEx++ \cite{Palangappa:2017:CCC:3058793.3050440} integrates custom expansion codes and variable compressibility. These methods reduce the total energy consumption of the memory system by almost half, along with providing improvements in IPC and bandwidth.

Along with the other issues that MLC cells face, another key issue is the high read latency. In most of the common architectures \cite{6477188, 4735567, Qureshi:2010:MMS:1816038.1815981}, reading the values of an MLC cell works in an iterative fashion, where the values of the MSB are detected, as the resistance values differ by a large amount. Following this, the further bits are detected, with smaller consecutive resistance gaps. This method resembles a binary search across the possible resistance values of the states. A solution \cite{Hoseinzadeh:2015:SSP:2836331.2829951} presented to reduce the required iterations for reads is by striping multiple lines and grouping them on a single array. This method improves the read latency and IPC, at a small cost of lifetime.

\begin{figure}[h]
    \centering
    \includegraphics[width=3in]{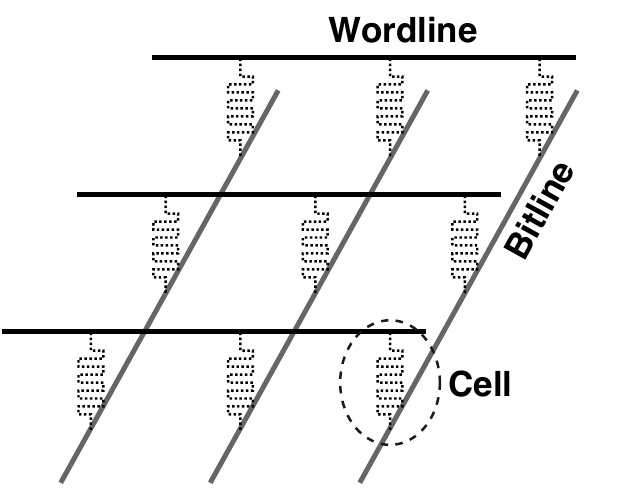}
    \hfil
    \caption{Crossbar structure of ReRAM cells.}
    \label{fig:ReRAM_Diagram}
\end{figure}

\subsection{Accelerators}
Research has shown that PCM \cite{7151827}, STT-RAM \cite{6865325, 7086084}, and ReRAM \cite{Hu:2012:HRB:2228360.2228448, Li:2013:MAC:2648668.2648729} are capable of performing operations and computations in addition to their capability of storing data. This method of having a device store data as well as perform computations allows for building in-memory accelerators which can quickly compute basic functions without requiring CPU intervention. Out of these ReRAM exhibits a crossbar array structure, as shown in Figure \ref{fig:ReRAM_Diagram}, that is beneficial for in-memory processing of matrixvector multiplication. Due to this, many accelerators exploiting the structure of ReRAM have arisen \cite{JOURTOP, Kim:2015:RDN:2767119.2700234, 7409722}. 

Chi et al. \cite{Chi:2016:PNP:3001136.3001140} used the structure in order to accelerate neural network computations. Shafiee et al. \cite{Shafiee:2016:ICN:3001136.3001139} focussed on applying ReRAM for Convolutional Neural Network (CNN) applications. However, both of these methodologies did not work well with training and weight updations for CNNs. In an attempt to rectify this, Song et al. \cite{7920854} built on these works, taking advantage of intra-layer parallelism to create an architecture that greatly boosts speed during both the training and inference phase of CNNs. 

Besides focusing on neural network computations, Bojnordi et al. \cite{7446049} explored the usage of ReRAM in developing an accelerator for Boltzmann machine, resulting in a large improvement in performance along with reduced energy consumption when compared with a multicore implementation. Song et al. \cite{8327035} proposed a structure to speed graph processing of graph algorithms that can be expressed as sparse matrix vector multiplication. Mittal compiled a survey \cite{make1010005} which explains neural network and processing in-memory accelerators that utilize ReRAM in more detail.

\begin{table*}
\begin{center}
\caption{Classification of Research Work}
\label{tbl:research}
\begin{tabular}{c >{\centering\arraybackslash}p{3in}}
\hline
\\
\textbf{Field of Work} & \textbf{Papers} \\
\\
\hline
\\
Lifetime Improvement & \cite{7459475, Garcia:2017:CLE:3132402.3132411, Zhang:2016:MWE:3001136.3001192, Palangappa:2017:CCC:3058793.3050440, 8342274, 7446056, 6880482, Qureshi:2009:ELS:1669112.1669117, 8203453, 7027229, 8611205, Chen:2013:OCB:2463585.2463592, Mittal:2014:WIL:2591513.2591525, 6522322} \\ \\
Error-Correction & \cite{7459409, 7927153, Tavana:2017:RRM:3132402.3132421, Swami:2017:EES:3154814.3151083, Tavana:2018:BCA:3274266.3243906, Schechter:2010:UEE:1816038.1815980, 7851482, Azevedo:2013:ZME:2508148.2485961, 5749752, 7847645, Ipek:2010:DRM:1735971.1736023}\\ \\
Performance and Parallelism & \cite{Qureshi:2012:PIP:2366231.2337203, 7459408, 7459410, 8342114, Li:2016:MAP:3012405.3012007, 7551433, 6522326, 5416645, Xia:2014:DDW:2597652.2597661, 6489977, 7059026, 7573814}\\ \\
Multi-Level Cells & \cite{7459508, 7920841, 8327021, Hoseinzadeh:2015:SSP:2836331.2829951, Rashidi:2018:IMP:3199680.3177965, Seong:2013:TPC:2485922.2485960, 6296504, 4735567, Qureshi:2010:MMS:1816038.1815981, 6853228, 6169027, 4418973, 5416645, 6241871, 5958219, Jiang:2012:FFP:2457472.2457481, 6168941, Li:2013:CDW:2499369.2465564}\\ \\
Accelerators & \cite{7459444,  7920854, 8327034, Chi:2016:PNP:3001136.3001140, Imani:2016:PAR:2989081.2989086, Fey:2016:UMT:2989081.2989124, 7927095, 7926952, Fujiki:2018:IDP:3173162.3173171, 8342237, 8342150, 8342118, 8342168, 8342186, 8342193, 8342229, Zhao:2018:LMS:3240302.3240422, 7446049, 8327035, Hu:2012:HRB:2228360.2228448, Li:2013:MAC:2648668.2648729, 6865325, 7086084, 7151827, JOURTOP, Kim:2015:RDN:2767119.2700234, 7409722, Shafiee:2016:ICN:3001136.3001139}\\ \\
\hline
\\
\textbf{SCM Type} & \textbf{Papers} \\
\\
\hline
\\
ReRAM & \cite{8327034, Chi:2016:PNP:3001136.3001140, Imani:2016:PAR:2989081.2989086, Fey:2016:UMT:2989081.2989124,
7927095, 7926952, Fujiki:2018:IDP:3173162.3173171, 8342237, 8342150,
8342118,8342168,
8342186,8342193,
8342229,7446049,
8327035,Hu:2012:HRB:2228360.2228448,
Li:2013:MAC:2648668.2648729,
JOUR,
Kim:2015:RDN:2767119.2700234,
7409722,
Shafiee:2016:ICN:3001136.3001139,
7927153,
Swami:2017:EES:3154814.3151083,
Tavana:2018:BCA:3274266.3243906,
Azevedo:2013:ZME:2508148.2485961,
5749752,
Zhang:2016:MWE:3001136.3001192,
Palangappa:2017:CCC:3058793.3050440,
7446056,
6880482,
8203453,
7027229,
8611205,
Mittal:2014:WIL:2591513.2591525,
6522322,
7459408,
Strukov2016,
6168941,
Ipek:2010:DRM:1735971.1736023}
\\ \\
STT-RAM & \cite{8327034,
Zhao:2018:LMS:3240302.3240422,
6865325,
7086084,
7459409,
Swami:2017:EES:3154814.3151083,
Tavana:2018:BCA:3274266.3243906,
Azevedo:2013:ZME:2508148.2485961,
5749752,
Palangappa:2017:CCC:3058793.3050440,
6880482,
8203453,
7027229,
8611205,
Chen:2013:OCB:2463585.2463592,
Mittal:2014:WIL:2591513.2591525,
6522322,
7459508,
8342114,
Strukov2016,
6168941,
Ipek:2010:DRM:1735971.1736023}
\\ \\
PCM & \cite{8327034,
7151827,
Tavana:2017:RRM:3132402.3132421,
Swami:2017:EES:3154814.3151083,
Tavana:2018:BCA:3274266.3243906,
Schechter:2010:UEE:1816038.1815980,
7851482,
Azevedo:2013:ZME:2508148.2485961,
5749752,
7847645,
7459475,
Garcia:2017:CLE:3132402.3132411,
Zhang:2016:MWE:3001136.3001192,
Palangappa:2017:CCC:3058793.3050440,
8342274,
7446056,
6880482,
Qureshi:2009:ELS:1669112.1669117,
8611205,
7920841,
8327021,
Hoseinzadeh:2015:SSP:2836331.2829951,
Rashidi:2018:IMP:3199680.3177965,
Seong:2013:TPC:2485922.2485960,
6296504,
4735567,
Qureshi:2010:MMS:1816038.1815981,
6853228,
6169027,
4418973,
5416645,
6241871,
5958219,
7459410,
Li:2016:MAP:3012405.3012007,
Qureshi:2012:PIP:2366231.2337203,
6522326,
Xia:2014:DDW:2597652.2597661,
6489977,
7059026,
6489977,
7573814,
doi:10.1063/1.3653279,
Strukov2016,
6477188,
Zhou:2009:DEE:1555815.1555759,
6176872,
Jiang:2012:FFP:2457472.2457481,
6168941,
Ipek:2010:DRM:1735971.1736023,
Li:2013:CDW:2499369.2465564,
1493077}
\\ \\
\hline
\end{tabular}
\end{center}
\end{table*}

\section{Simulators} \label{Simulators}
While several options for system simulation exist, the most popular simulators for SCM are GEM5 \cite{Binkert:2011:GS:2024716.2024718}, NVSim \cite{6218223} and NVMain \cite{6296505, 7038174}. 

GEM5 is a simulator commonly used in computer architecture and system research. It can simulate system-level and processor microarchitecture. Provisions exist for full-system capabilities of Alpha, ARM, SPARC, and x86 simulations. The CPU model being simulated can be interchanged between a simple, an inorder, or a full-scale out-of-order CPU. For out-of-order CPU, traces can be run to obtain detailed results on the performance of the memory system. 

NVSim is a circuit-level modelling tool, which provides estimations for the performance, energy, and area values for a given design specification. It supports the commonly used NVM memories, such as PCM, STT-RAM, ReRAM, as well as NAND based Flash memory. The intent of the tool is to help in creating optimized designs for the metrics mentioned, before fabricating the physical chip. NVSim is based off of the existing analytical model, CACTI \cite{509850, thoziyoor2008cacti}.

NVMain is a main memory simulator targeted for use with NVM-based memories. NVMain provides a cycle-accurate simulator which can estimate energy consumption at the system level. NVMain supports main memories of DRAM, PCM, STT-RAM, and ReRAM, as well as hybrid versions of these memories. NVMain has the option to be used in conjunction with GEM5 in order to evaluate full system simulation.


\section{Future Outlook} \label{Future}
Due to the many shortcomings of a pure SCM-based memory system, the majority of research now deals with hybrid memory. This gives the benefit of scalability, while allowing the shortcomings of SCM to be offset by a DRAM/SRAM component. With the release of Intel's 3D-XPoint memory \cite{newsroomintel} the use of this memory has gone from being hypothetical to practical, with claims that it performs 1000 times faster than NAND SSDs, with 1000 times the endurance. Instead of replacing existing layers of memory as was previously expected, current NVM memory is trying to bridge the gap between long latency NAND-based secondary memory, and  primary memory, providing a large capacity of storage in doing so. In this way, NVM is seeking to change the memory hierarchy from a set of discrete memory layers to a spectrum of different memory possibilities. 

Another aspect of SCM being actively looked at is its unique capability of supporting MLC. As mentioned previously, the use of MLC comes a new range of problems that have to be overcome before widespread use becomes practical. The International Technology Roadmap for Semiconductors (ITRS) \cite{Allan:2002:TRS:619071.621883,
MARDEN2002371} predicts that densities of 3-bit and 4-bit MLCs may become possible in the near future. This increase in density may further exacerbate the existing problems that are being experienced.

\addtolength{\textheight}{0cm}   


\bibliographystyle{IEEEtran}
\bibliography{references}
\end{document}